\documentstyle[aps,prb,preprint,graphicx]{revtex}

\newcommand{\etal}{\emph{et al. }}
\newcommand{\Hamil}{{\cal H}}
\newcommand{\ket}[1]{|{#1}\rangle}
\newcommand{\bra}[1]{\langle{#1}|}

\newcommand{\eqgr}[2]{(-i)\langle {\rm T}{#1}{#2}\rangle}

\newcommand{\av}[1]{\langle{#1}\rangle}

\newcommand{\anticom}[2]{\{{#1},{#2}\}}

\newcommand{\ddtinline}{\partial/\partial t}

\newcommand{\ov}[2]{{\cal O}_{{#1}{#2}}}

\newcommand{\ovl}[2]{{\cal O}^{-1}_{{#1}{#2}}}

\def\ket#1{|#1\rangle}
\def\bra#1{\langle#1|}
\def\inner#1#2{\mathop{\langle#1|#2\rangle}}

\def\bfr{{\bf r}}

\def\dmu#1{{\rm d}#1}

\def\leade#1{\varepsilon_{#1\sigma}}
\def\dote#1{\varepsilon_{#1}}

\def\c#1{c_{#1\sigma}}
\def\cdagger#1{c_{#1\sigma}^{\dagger}}

\begin{document}
%\preprint{\today}
\title{Effects of non-orthogonality in the time-dependent current through tunnel 
junctions}
\author{J. Fransson$^1$, O. Eriksson$^1$ and I. Sandalov$^{1,2,3}$}
\address{$^1$ Condensed Matter Theory group, Uppsala University, Box 530, 751 21\ \ Uppsala, Sweden\\
$^2$ Department of Physics, Link\"oping University, 581 83\ \
Link\"oping, Sweden\\ $^3$ Kirensky Institute of Physics, RAS, 660036
Krasnoyarsk, Russian Federation}
\maketitle

\begin{abstract}
A theoretical technique which allows to include contributions from
non-orthogonality of the electron states in the leads connected to a
tunneling junction is derived. The theory is applied to a single
barrier tunneling structure and a simple expression for the
time-dependent tunneling current is derived showing explicit
dependence of the overlap. The overlap proves to be necessary for a
better quantitative description of the tunneling current, and our
theory reproduces experimental results substantially better compared
to standard approaches.
\vspace{2mm}
\end{abstract}

Achievements in nano-materials science is expected to have importance
in many scientific fields, including information technology, quantum
computing and fuel cells. In particular, tunneling phenomena have
been under focus recently, both in magnetic heterostructures as well
as for quantum dot systems. The purpose of this paper is to develop an
improved description of this phenomenon for general tunnel junctions,
with possible application to the aforementioned scientific questions.

To focus the discussion, we mention that conductance measurements on
extremely small metal-insulator-metal (MIM) junctions were carried out
by Vullers \emph{et al.}\cite{vullers2000} showing a non-linear
conductance as a function of the bias voltage for low
temperatures. The same behaviour has been reported for MIM double
junctions\cite{nakayama1999} and Ti/TiO$_x$ tunneling barrier
systems\cite{irmer1997,matsumoto1996,haraichi1997}. The non-linearity
in the current-voltage $(J-V)$ characteristics appears for
source-drain bias voltages larger than the spacing of the quasi one
dimensional subbands since different numbers of subbands become
available for transport in the forward and reverse directions
\cite{glazman1989}. In the study by Simmons \cite{simmons1963} the
current was found to depend non-linearly on the voltage, roughly as
$V+\gamma V^3$.

Many theoretical studies of transport in nanostructures with tunneling
barriers rely on the transfer
Hamiltonian\cite{bardeen1961,payne1986,larkin1987,meir1992,jauho1994}
which contains serious inconsistencies \cite{svidsinskii1982}. The
principle of the transfer Hamiltonian is a division of the system into
subsystems. This is motivated by the fact that the physical properties
of the subsystems may be different and, hence, require different
descriptions. Another motivation is that one is directly offered the
possibility to generalize the approach to any number of tunneling
barriers in the system.  The transfer (tunneling) between the
subsystems arises due to an overlap of the wave functions in the
region of the barrier whereas the electron operators of the different
subsystems are assumed to be anti-commuting. Qualitatively this may be
motivated since the leakage of a wave function in one subsystem into
the other is exponentially small. The $J-V$ characteristics given in
this picture also shows a non-linear structure for large bias
voltages. Quantitatively, though, the assumption of anti-commuting
operators creates serious errors in the calculations of the
current. This becomes particularly evident in the equilibrium
situation displayed in Table \ref{table-equilibrium}, in which the
four lowest states of a particle in a one dimensional hard walled box
with a scattering potential are given. The energy levels are, as
expected, reproduced within the non-orthogonal representation (NOR)
with much higher accuracy than in the orthogonal representation
(OR). Attempts that go beyond the transfer Hamiltonian have been made,
\emph{e.g.} by expanding the non-orthogonal states into a new Hilbert
space \cite{emberly1998}. However, the proven success and physical
transparency of the transfer Hamiltonian approach makes it desirable
to extend its applicability to more general situations where the
overlap is large, without making use of perturbation theory. This can
indeed be achieved, which we demonstrate in this paper.

In order to overcome the inconsistencies with the transfer Hamiltonian
formalism, we develop a theoretical approach for time-dependent
transport through tunneling systems in which the overlap between the
subsystems give explicit contribution to the current. Technically, we
will express the properties of the original system in terms of the
operators constructed of the wave functions of each subsystem. The
resulting model structurally resembles the transfer Hamiltonian,
although the physical interpretation is different. We have chosen the
single barrier system simply to show the features of our approach. The
main result of this paper is the Eqn. (\ref{eq-JIcurrent}) for the
time-dependent tunneling current through a single barrier. This
expression is applied to a MIM junction in order to analyze the effect
of overlap on the current. To our knowledge there does not exist any
derivation nor analysis of time-dependent transport in tunneling
junctions where the non-orthogonality is \emph{not} disregarded.

Let us now proceed  starting with the one particle Hamiltonian
\[ H=\frac{p^2}{2}+V,
\]
where $V$ is any potential describing a system of two leads with an
insulating layer in between. We introduce the two potentials
$V_{\alpha},\ \alpha=L,R$, for the left $(L)$ and the right $(R)$
subsystem, respectively\cite{bardeen1961,payne1986,fran1999}. For
instance, the left potential can be written as
$V_L=V(x)\theta(-x+a_L)+V(a_L)\theta(x-a_L)$, where $\theta(x)$ is the
Heaviside function and $a_L$ is a turning point for the left
subsystem. In each subsystem there are orthonormal eigenstates
$\{\phi_k,\leade{k}\}_{k\sigma\in\alpha}$ from which the corresponding
field operator
$\psi_{\alpha}(t,x)=\sum_{k\sigma\in\alpha}\c{k}(t)\phi_k(x)$ is
constructed. Here $t$ is time and $x=(\bfr,\sigma)$ is a vector of the
spatial coordinate $\bfr$ and the spin $\sigma$.  Suppose that $\psi$
is the field operator of the system formed by the potential $V$. Then,
this operator can be expanded in terms of $\psi_{\alpha}$ by the
trivial identity
$\psi(t,x)=\sum_{\alpha}\psi_{\alpha}(t,x)+[\psi(t,x)-\sum_{\alpha}\psi_{\alpha}(t,x)]$. Following
reference\cite{sandalov1990} we project $\psi$ onto the subsystem
$\alpha$ by
\[ \tilde{c}_{k\sigma}(t)=\int\phi_k^*(x)\psi(t,x)\dmu{x},
\]
$k\in\alpha$, interpreted as the annihilation of a particle in the
state $\phi_k$ with spin projection $\sigma$. Creation
$\tilde{c}^{\dagger}_{k\sigma}$ of a particle in the state $\phi_k$ is
defined similarly. These projections are possible to use directly for
a second quantized form of the Hamiltonian. However, such an expansion
gives an inconvenient expression of the Hamiltonian with the overlap
matrix appearing explicitly. Thus, in order to proceed further, we
define the operators
\begin{equation}
\begin{array}{rcl}
\c{k}(t)&=&\sum_{k'}\ovl{k}{k'}\tilde{c}_{k'\sigma}(t),
\\
\cdagger{k}(t)&=&\sum_{k'}\left(\ovl{k}{k'}\right)^*
			\tilde{c}^{\dagger}_{k'\sigma}(t),
\end{array}
\label{eq-annihilator}
\end{equation}
where $k'$ runs over all states in $L\cup R$ and $\ovl{k}{k'}$ is the
element $kk'$ of the inverse of the overlap matrix of the wave
functions $\phi_k,\ \phi_{k'}$ given by
$\ov{k}{k'}=\inner{\phi_k}{\phi_{k'}}=\ov{k'}{k}^*$. By a limitation
to the case of spin conservation we can omit the spin indices in the
overlap integral. The expectation value of the Hamiltonian in these
operators is
\begin{equation}
\Hamil=\int\psi^{\dagger}H\psi\dmu{x}=
	\Hamil_L+\Hamil_R+\Hamil_T,
\label{eq-hamiltonian1}
\end{equation}
where we have defined
$\Hamil_{\alpha}=\int\psi_{\alpha}^{\dagger}H\psi_{\alpha}\dmu{x}$ and
$\Hamil_T=\sum_{\alpha\alpha'}(\int\psi_{\alpha}^{\dagger}H\psi_{\alpha'}\dmu{x}+H.c.)$. Here,
we have neglected all expectation values that contain
$\psi-\sum_{\alpha}\psi_{\alpha}$. Furthermore, we note that from the
identity $V=V_{\alpha}+[V-V_{\alpha}],\
\alpha=L,R$, we find that the Hamiltonian of the lead
$\Hamil_{\alpha}=\sum_{k\sigma\in\alpha}\leade{k}\cdagger{k}\c{k}+\sum_{kk'\in\alpha}\bra{\phi_k}(V-V_{\alpha})\ket{\phi_{k'}}\cdagger{k}\c{k'},\
\alpha=L,R$. The last term is a sum of terms proportional to the
integral of $\phi_k^*\phi_{k'}$ over $(a_R,\infty)$ or $(-\infty,a_L)$
when $\alpha=L$ or $\alpha=R$, respectively, in which domains the wave
functions are exponentially small. Thus, this term is negligible and
we arrive at the appealing form of the Hamiltonian
\begin{eqnarray}
\Hamil=	\sum_{p\sigma\in L}\leade{p}\cdagger{p}\c{p}+
		\sum_{q\sigma\in R}\leade{q}\cdagger{q}\c{q}
	+\sum_{pq\sigma}(v_{pq\sigma}\cdagger{p}\c{q}+H.c.),
\label{eq-hamiltonian}
\end{eqnarray}
where $v_{pq\sigma}=\bra{\phi_p}H\ket{\phi_q}$ is the mixing matrix
element. The structure of the Hamiltonian (\ref{eq-hamiltonian}) very
much resembles the usual transfer Hamiltonian. Nevertheless, the
meaning of the electron operators $\cdagger{k},\c{k}$ is altered, now
carrying information of the full system rather than just of its
subsystem. This fact is legible from the anti-commutation relation
$\anticom{\c{k}}{\cdagger{k'}}=\ovl{k}{k'}$. Indeed, when
$\ovl{k}{k'}\rightarrow\delta_{kk'}$ we recover the transfer
Hamiltonian with the usual interpretation of the operators $\c{k}$. In
this sense we conclude that the Eq. (\ref{eq-hamiltonian}) generalizes
the conventional transfer Hamiltonian.

The expression in Eqn. (\ref{eq-hamiltonian}) is derived for the
system in equilibrium. It is straight forward applicable to the
non-equilibrium case by letting $\leade{k}\rightarrow\leade{k}(t)$ and
$v_{pq\sigma}\rightarrow v_{pq\sigma}(t)$. For definiteness we derive
an expression for the current flowing through the barrier from the
left to right. The tunneling current through the barrier separating
the leads is expressed as the rate of change of the number of
particles on, say, the left side of the junction
$\av{N_L(t)}=\sum_{p\sigma}\av{n_{p\sigma}(t)}$, where
$\av{n_{p\sigma}(t)}=\av{\cdagger{p}(t)\c{p}(t)}$. The time
development of $\av{n_{p\sigma}}$ is given by the Heisenberg equation
of motion yielding the tunneling current for each spin projection
$\sigma$
\begin{eqnarray}
J_{\sigma}(t)&=&
2e\Im\sum_{pq}\biggl(
	V^*_{pq\sigma}(t)\av{\cdagger{q}(t)\c{p}(t)}+
	v^*_{pq\sigma}(t)\ovl{p}{q}\av{\cdagger{p}(t)\c{p}(t)}
	\biggr)=
\nonumber\\
&=&-2e\Re\sum_{pq}\biggl(
	V^*_{pq\sigma}(t)F^<_{pq\sigma}(t,t)-
	v^*_{pq\sigma}(t)\ovl{p}{q}g^<_{p\sigma}(t,t)\biggr),
\label{eq-dynamicnumber}
\end{eqnarray} 
with the coefficients $V_{pq\sigma}=v_{pq\sigma}+\ovl{p}{q}\leade{q}$
describing the tunneling. In Eq. (\ref{eq-dynamicnumber}) we have
identified the correlation function $\av{\cdagger{q}\c{p}}$ with the
\emph{lesser} Green function
$F^<_{pq\sigma}(t,t)=i\av{\cdagger{q}(t)\c{p}(t)}$.  This propagator
is calculated within the non-equilibrium technique of Kadanoff and
Baym \cite{kada62} for the Green function
$F_{pq\sigma}(t,t')=\eqgr{\c{p}(t)}{\cdagger{q}(t')}$. From the
equation of motion for $F_{pq\sigma}(t,t')$ we obtain
\begin{eqnarray}
 F_{pq\sigma}(t,t')&=&g_{p\sigma}(t,t')\ovl{p}{q}+
\int_0^{-i\beta}g_{p\sigma}(t,t_1)
		V_{pq\sigma}(t_1)g_{q\sigma}(t_1,t')\dmu{t_1},
\label{eq-transferGF}
\end{eqnarray}
where $g_{k\sigma}=F_{kk\sigma}$ is the conduction electron Green
function satisfying the equation
$(i\ddtinline-\leade{k})g_{k\sigma}(t,t')=\delta(t-t')$. The contour
integration in Eq. (\ref{eq-transferGF}) is brought to real time
integration by the Langreth analytical continuation rules
\cite{langreth1976}, thus
\begin{eqnarray*}
F_{pq\sigma}^<(t,t')=g^<_{p\sigma}(t,t')\ovl{p}{q}+
	\int_{-\infty}^{\infty}V_{pq\sigma}(t_1)
	\biggl[g_{p\sigma}^r(t,t_1)g_{q\sigma}^<(t_1,t')+
		g_{p\sigma}^<(t,t_1)g_{q\sigma}^a(t_1,t')\biggr]\dmu{t_1}.
\end{eqnarray*}
The lesser, retarded and advanced expressions of the conduction
electron GF are
\begin{eqnarray*}
g^<_{k\sigma}(t,t')&=&if_{\alpha}(\leade{k})
		e^{-i\int_{t'}^t\leade{k}(t_1)\dmu{t_1}},\\
g^{r,a}_{k\sigma}(t,t')&=&
		\mp i\theta(\pm t\mp t')e^{-i\int_{t'}^t\leade{k}(t_1)\dmu{t_1}},
\end{eqnarray*}
respectively, where $f_{\alpha}(x)$ is the Fermi-Dirac distribution
function. Before we continue the derivation we rewrite the electron
operators in terms of current states, \emph{i.e.}
$\cdagger{k}(t)=\cdagger{k}\exp{[i\mu_{\alpha}(t)]}$, and
$\c{k}(t)=\c{k}\exp{[-i\mu_{\alpha}(t)]}$. This will explicitly show
the applied voltage dependence $V(t)$ of the current, since
$\mu_L(t)-\mu_R(t)=eV(t)$. Replacing the summation over $p$ and $q$ in
Eq. (\ref{eq-dynamicnumber}) by energy integration in terms of the
density of states $\rho_{\sigma}(\dote{\alpha})$ and noting that
$\Re[(V_{pq\sigma}^*-v_{pq\sigma}^*)\ovl{p}{q}g^<_{p\sigma}]=0$, the
time-dependent tunneling current becomes
\begin{eqnarray}
J_{\sigma}(t)&=&-2e\Re\int V_{LR\sigma}^*(t)
	\rho_{\sigma}(\dote{L})\rho_{\sigma}(\dote{R})
\nonumber\\
&&	\times\int_{-\infty}^t 
	V_{LR\sigma}(t_1)[f(\dote{R})-f(\dote{L})]	
	e^{-i\int_{t_1}^t(eV(t_2)+(\dote{L}-\dote{R}))\dmu{t_2}}
	\dmu{t_1}\dmu{\dote{L}}\dmu{\dote{R}}.
\label{eq-JIcurrent}
\end{eqnarray}
The mixing and the overlap are here replaced by the functions
$V_{LR\sigma}(t)\equiv V_{\sigma}(\dote{L},\dote{R},t)$ and
$\ovl{L}{R}\equiv\ovl{}{}(\dote{L},\dote{R})$, respectively,
satisfying $V_{\sigma}(\leade{p},\leade{q},t)=V_{pq\sigma}(t)$ and
$\ovl{}{}(\leade{p},\leade{q})=\ovl{p}{q}$.  The formula
(\ref{eq-JIcurrent}) reproduces results based on the transfer
Hamiltonian in the limit of orthogonal subsystems, \emph{i.e.} when
$\ovl{k}{k'}\rightarrow\delta_{kk'}$. It is important to note the fact
that the tunneling coefficient
$V_{LR\sigma}=v_{LR\sigma}+\ovl{L}{R}\leade{R}$ in our formulation,
explicitly depends on the energies of the electrons involved in the
conduction process.

\begin{figure}[h]
\begin{center}
\includegraphics[width=8cm]{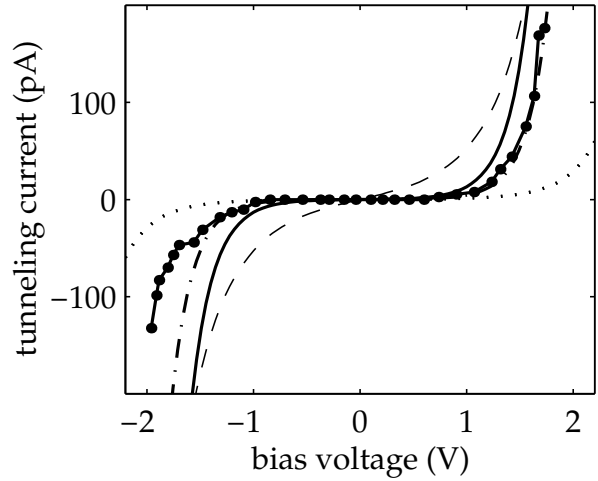}
\end{center}
\caption{The $J-V$ characteristics of a 1.46 nm wide and 1.85 eV high MIM junction (the height measured from the equilibrium chemical potential) [19]. The experimental results by Haraichi \emph{et al.} [5] (solid-dotted) is compared with the computations within the NOR (solid), NOR with a 6 $\%$ increase of the width (dash-dotted), OR (dashed) and Simmons formula (dotted) [7]. The equilibrium chemical potential is 1.75 eV and the conduction band width is $2W=40$ eV.}
\label{fig-JIstationarycurrent}
\end{figure}

When $V(t)=V$ and a stationary current is established through the
barrier the Eq. (\ref{eq-JIcurrent}) reduces to
\begin{eqnarray}
J_{\sigma}=2e\frac{\pi}{4W^2}\int_{-W}^{W}
	|V_{LR\sigma}|^2[f(\dote{}-eV)-f(\dote{})]
	\dmu{\dote{}}.
\label{eq-JIstationarycurrent}
\end{eqnarray}
This expression is given by assuming a constant density of states
$\rho_{\sigma}(\dote{\alpha})=1/2W$, where $2W$ is the conduction band
width, and slowly varying mixing and overlap so that their respective
values can be taken at the chemical potential, which are reasonable
conditions for MIM-junctions. In order to compare our theory with a
realistic example we show in Fig. \ref{fig-JIstationarycurrent} the
experimental $J-V$ characteristics from Ref. \cite{haraichi1997}
(solid-dotted line) together with that of
Eqn. (\ref{eq-JIstationarycurrent}) in both the non-orthogonal (solid
line) and orthogonal (dashed line) representations. We have also
included the corresponding result given by Simmons formula (dotted
line) \cite{simmons1963}. Note that Simmons formula and the orthogonal
representation correspond to the standard methods used to calculate
transport. From the figure, it stands clear that inclusion of the
overlap contributes significantly to the behaviour of the $J-V$
characteristics and the quantitative agreement is remarkably
improved. The increase in the agreement with the experiments lies not
only in the low voltage regime but also in that the current rises
rapidly at a certain threshold voltage, which influences the
time-dependent current. For a $6\ \%$ increase in the barrier width
our calculation (bold dash-dotted line) agrees exactly with the
experimental results for positive voltages. The remaining discrepancy
from the experimental curve, \emph{e.g.} the observed asymmetry, is
believed to stem from the lack of electron interactions in our model,
for example charging effects. Moreover, in the simple calculations
presented here we have merely computed the wave functions
$\phi_{\mu_L}$ and $\phi_{\mu_R}$, normalized to a unit probability
flow \cite{landau1977} at their asymptotic distances from the barrier
$x\rightarrow-\infty$ and $x\rightarrow\infty$. For simplicity we have
used a rectangular potential barrier leading.

In conclusion, we have developed a simple and transparent theoretical
approach for time-dependent tunneling current through nanostructures
which has a far wider applicability compared to standard methods. The
ability of dividing the system into several subsystems, which then can
be treated individually, is preserved without loss of accuracy when
the inclusion of the overlap of the subsystems is allowed and all
attractive features of the transfer Hamiltonian approach can be
kept. The non-orthogonality is reflected in the non-zero
anti-commutation relations of the electron operators of different
subsystems. A formula for the time-dependent tunneling current through
a single barrier structure, Eqn. (\ref{eq-JIcurrent}), has been
derived, which shows the necessity of including the overlap for a
substantially better quantitative agreement with experiments. We also
note that the formalism simply generalizes to the case of a two, or
multiple, barrier structure. In particular, the region between the
barriers can be interacting, for example a quantum dot. Then, a
generalization to any number of contact leads is straight forward.

J.F. wants to thank U. Lundin for helpful and encouraging
discussions. Support from the G\"oran Gustafsson foundation, the
Swedish national science foundation (NFR and TRF) and the Swedish
foundation for strategic research (SSF) are acknowledged.

\newpage
\begin{table}[h]
\caption{The four lowest energy levels of a 37 nm long hard walled box with a 5.3 nm wide and 178 meV high scattering potential located in the middle of the box. The energies (meV) are computed exact, with the overlap matrix taken into account (NOR) and ignored (OR).}
\begin{center}
\begin{tabular}{crrrc}
&exact & NOR & OR\\\hline &20.265 & 20.266 & 18.866 \\ &27.781 &
27.862 & 27.342 \\ &83.868 & 83.592 & 79.383 \\ &111.088 & 113.793 &
107.176
\end{tabular}
\end{center}
\label{table-equilibrium}
\end{table}

\newpage

\end{document}